\documentclass{article}
\usepackage{times,uweetr}

\usepackage{cite}

   \usepackage[dvips]{graphicx}
   \DeclareGraphicsExtensions{.eps}

\usepackage[cmex10]{amsmath}

\usepackage{subfig}
\usepackage{fancyvrb}
\usepackage{verbatim}
\usepackage{moreverb}

\begin{document}

\title{Analysis of Internet Measurement Systems for Optimized Anomaly Detection System Design}
\author{Sean McPherson\and Antonio Ortega\thanks{This work is supported in part by the National Science Foundation's Networking Technology and Systems (NeTS) program, grant number CNS-0626696.}\\
{\tt\ smcphers@usc.edu,ortega@sipi.usc.edu}\\
\\
University of Southern California\\
Signal and Image Processing Institute\\
Ming Hsieh Department of Electrical Engineering, 3740 McClintock Avenue,\\ Los Angeles, CA 90089-2565\\
}

\reportmonth{July}
\reportyear{2009}
\reportnumber{0000}

\maketitle

\begin{abstract}
Although there exist very accurate hardware systems for measuring traffic on the internet, their widespread use for analysis tasks is limited by their high cost. On the other hand, less expensive, software-based systems exist that are widely available and can be used to perform a number of simple analysis tasks. The caveat with using such software systems is that application of standard analysis methods cannot proceed blindly because inherent distortions exist in the measurements obtained from software systems. The goal of this paper is to analyze common Internet measurement systems to discover the effect of these distortions on common analysis tasks. Then by selecting one specific task, periodic signal detection, a more in-depth analysis is conducted which derives a signal representation to capture the salient features of the measurement and develops a periodic detection mechanism designed for the measurement system which outperforms an existing detection method not optimized for the measurement system. Finally, through experiments the importance of understanding the relationship between the input traffic, measurement system configuration and detection method performance is emphasized.
\end{abstract}

\section{Introduction}
\label{sec:intro}
Practical internet network traffic measurement systems have been available for a number of years and have led to the development of a variety of analysis tools, such as bandwidth estimation techniques, to extract useful information from measurements or for network security purposes like detecting maltraffic traversing the network. Typically, selecting an Internet measurement system tends to be based on characteristics such as cost or maximum time resolution, e.g., if the application demands high timing precision one may select a higher cost hardware measurement system. We argue that measurement system selection should also consider the specific application, along with related data representation and data processing issues. Understanding how a measurement system affects the precision of data being collected makes it possible to improve the application performance by
selecting appropriate data representation and tools.  As an example application, we consider the problem of detecting periodic denial-of-service (DOS)
attacks, which can be characterized by the presence of a relatively low rate periodic traffic in otherwise ``random'' aggregate traffic. Clearly this problem could be approached by selecting a very high resolution measurement system (e.g., a hardware based one), but at a potentially prohibitive cost. A key contribution of our work is to show that in many cases lower end (e.g., software based) measurement systems can be used as effectively, {\em as long as the effect of the measurement system has been taken into account in designing the analysis techniques.}

Studying the impact of a measurement system on Internet measurement analysis applications, such as DOS detection, can sometimes be framed using standard digital signal processing (DSP) tools. However, a key difficulty is that most DSP based analyses are based on the assumption that the measurement timing (e.g., when samples are captured) is independent of the signal being measured. Instead, practical Internet measurement systems provide measurements that are {\em triggered by the signal being measured, i.e., by packet arrivals.} This is not a problem if individual packet 
arrival events are recorded separately and with a high precision clock, but it makes modeling the measurement system more difficult when 
timing is no longer reliable (because a host system has to be interrupted to record a packet arrival) and/or multiple packet arrival intervals are recorded jointly (due to so called interrupt coalescence). 

Most currently used analysis methods tend to ignore the timing distortion introduced by the measurement system and create a signal representing
arrivals per measurement interval. ``Off the shelf'' DSP tools are then used to analyze these signals, implicitly assuming that those measurement intervals are equal, which is not true in practice for typical Internet measurement systems.  Instead, we propose to use a detailed understanding of both measurement system and signal to be detected in order to design specific detection strategies. These strategies can then handle situations of high variability in measurement intervals. Towards this goal, we study measurement systems based on interrupt coalescence and propose a new detection system that can operate under challenging interrupt coalescence conditions.

This paper is organized as follows. Section~\ref{sec:measure} includes a general overview for both software and hardware network measurement systems describing the systematic differences and limitations inherent to software based systems. In Section~\ref{sec:analysis} we narrow our focus to software measurement systems using one type of interrupt coalescence. Here we define a suitable signal representation and perform an in-depth analysis of the measurement system effects for an idealized Poisson input signal. Following this analysis, in Section~\ref{sec:PAD} we attempt to detect a periodic signal using a detection method which expects an accurate, uniformly sampled signal and observe the loss in performance when, due to the measurement system, this is not the case. In Section~\ref{sec:chi} we discuss the intuition we used while designing our detection algorithm, and give a detailed description of its operation. Section~\ref{sec:experiment} provides an experimental comparison between our detection scheme and the previously mentioned detection method and provides some guidelines as to which detection method is the best choice given a measurement system and knowledge of traffic rates. 

\subsection{Related Work}
\label{sec:related}
Studying the effect of the measurement system is not an entirely new problem. Partridge \emph{et al.} examined the effect of signal representation on wireless network data traffic, and showed how a given representation could lead to different analysis methods in \cite{Cousins02usingsignal}. The difference in their work and ours is that they begin with accurate measurements and use different representations based on the analysis task. We on the other hand chose the representation that fits the measurement system most naturally and design analysis methods to fit the representation. 

Prasad \emph{et al.} looked at how interrupt coalescence altered standard network tomography methods, and discussed some methods to detect the presence of coalescence in measured signals in \cite{intcoalescence}. Our work is different because we do not simply show the effect of the measurement system, but instead we use our knowledge of the measurement system and its inherent effect to design strategies for a specific networking problem, namely DOS attack detection. We chose DOS attack detection because recent work \cite{Hussain06a,Mitra06a} has shown malware and specifically DOS attacks can produce periodic traffic. Thus DOS attack detection becomes a problem of periodic signal detection, which is a common signal processing task. For comparison we use an attack detection method, called the periodic attack detector (PAD), proposed by He in \cite{xinming}, which is was shown to produce good detection performance in \cite{Thatte08a}.  The PAD is different from our method because it is not designed for a specific measurement system.

In this paper we focus on periodic DOS attack detection, but our detection method could be used to detect the presence of periodic traffic in aggregate network streams due to other network anomalies. For instance, in \cite{DKatabi, PHuang} periodic traffic was shown to arise as a consequence of bottlenecks or similar network congestion problems. Because of this periodic nature, signal processing techniques (e.g., spectral analysis techniques based on wavelets) were applied to network measurements as a way to quickly detect underperforming networks.

\section{Measurement System Overview}
\label{sec:measure}
\begin{figure}[htb]	
\centering
\includegraphics[height=2in]{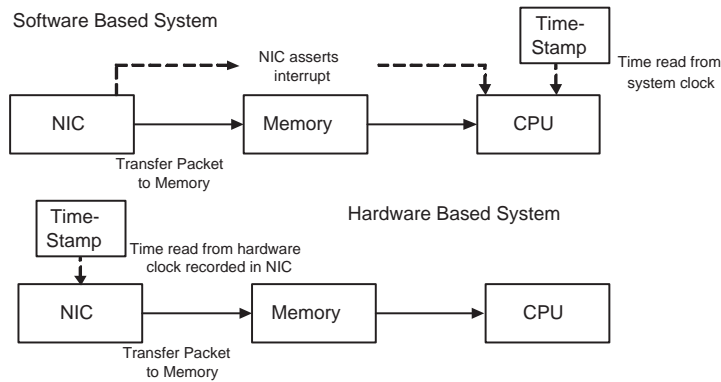}
	\label{fig:hardwaresoftware}
	\caption{System overview for both the software and hardware based network measurement system.}
\end{figure}

Measurement systems fall into two main categories: hardware and software based systems. As illustrated in Figure~\ref{fig:hardwaresoftware}
the key differentiating factor between hardware and software systems is how time stamps are generated for the incoming packets. In hardware based systems, like the popular DAG cards developed by Donnelly \cite{highquality}, the packet time stamp is generated by a very accurate hardware clock separate from the host system. The precision in the packet time stamps makes hardware system measurements amenable to standard DSP analysis tools.  Alternatively, in software based systems time stamps are generated in software, e.g., as in the Tstat tool \cite{tstat}, by querying the host system clock. Because the hardware based system time stamps the data before it is processed by the network interface card (NIC) many of the timing errors introduced in the NIC are avoided. We now focus on software based systems, discuss the possible timing errors associated with the system and describe how we approximate the effects of the timing errors in simulations which are conducted later. 

\subsection{Software Based Measurement Systems}
\begin{figure}[htb]	
\centering
\includegraphics[width=.75\textwidth]{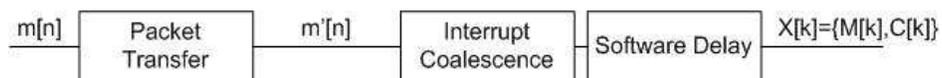}
	\label{fig:measurementsys3}
	\caption{Processing Delays in Software Measurement System}
\end{figure}

As mentioned above the key difference between hardware and software measurement systems how and when the incoming packets are time stamped. While a hardware system generates time stamps before it is processed by the NIC, in a software-based system the processing of the packet done by the NIC produces three timing delays, pictured in Figure~\ref{fig:measurementsys3}. The first source of delay is a packet size dependant delay that occurs in transferring the packet data to memory. The second delay is due to interrupt coalescence, which depends on the arrival timing of the incoming packet stream. Interrupt coalescence is more complicated than the data transfer delay, and can produce different types of error depending on the user configuration. The last form of delay occurs due to software, which queries the system clock for each received packet and creates artificial timing delays when multiple packet are received simultaneously (due to interrupt coalescence). Since these delays are packet size, system configuration and incoming packet rate dependent their effects can not, in general, be modeled or removed and therefore we do not have accurate timing in the measurements as was the case in a hardware system. Let $m[n]$ represent the time stamps of the packets as they are received at the NIC. Then we define $m'[n]$ to be the packet time stamp after the packet has been transferred to memory, and $X[k]$ is the vector representation (defined below) of the measurements following interrupt coalescence and software time stamping.

We now describe in detail the sources of delay associated with the software measurement system, and describe methods used to approximate the timing effects in our simulations. The analysis presented here is similar to the work of Salah and El-Badawi in \cite{livelockmodel} and \cite{infosciences}, however in their work they modeled the system components using queuing system models more common to networking. In our work we attempt to approximate the measurement system components using simple signal processing models, such as a delay, or by deriving an alternate signal representation which simplifies analysis yet still maintains the overall effect of the measurement system.


\subsubsection{Packet Transfer to Memory}
In a software based system the incoming packet data is first transferred to system memory. The NIC waits for this transfer to complete before asserting a system interrupt letting the CPU know that the packet data is available for processing. Commonly two methods are possible to transfer data from the NIC to memory, direct memory access (DMA) and programmed input output (PIO). With PIO the CPU is used to transfer the data from the NIC to memory. This often overburdens the CPU, which is already stressed in a high speed network measurement system, so in this paper we limit ourselves to the study of DMA data transfers \cite{infosciences}. 

The delay incurred in transferring the packet data to memory is a combination of two delays. The first is the time spent waiting to access the data transfer bus, which in this work we assume to be negligible. The second form of delay is the length of time required to actually transfer the data to memory. Because the speed of the DMA channel is fixed, the amount of time required to transfer a packet depends only on the size of the packet itself.

In this paper we assume that the time required to transfer the packet to memory is a constant multiple of the packet size, and we approximate this delay by simply adding $PacketSize/BitRate$ to the input packet time stamp to generate the new packet time stamp following data transfer. $BitRate$ is the bit rate of the network link and $PacketSize$ is the size in bits of the current packet. Thus our approximation represents the timing of the last bit of the packet, which reflects the packet size dependence of the data transfer component in the software measurement system. 

\begin{figure}[htb]	
\centering
\includegraphics[height=1.5in]{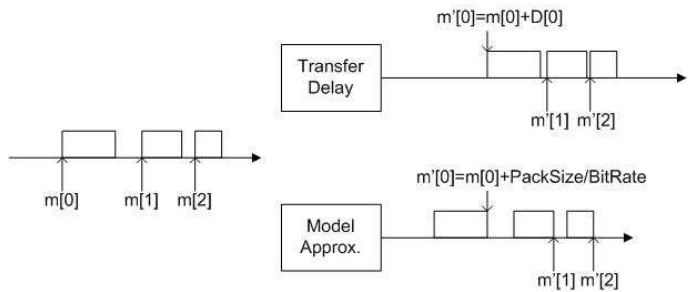}
	\label{fig:transfermodel}
	\caption{Example of Packet Size Dependant Transfer Model vs. Approximation of Model}
\end{figure}

Figure\protect~\ref{fig:transfermodel} shows an example of the packet size dependant data transfer model along with our approximation of the model. As defined above $m[n]$ represents the time stamp of the beginning of an packet received at the NIC. The delay $D[n]$ is the delay incurred in transferring the packet data to the host system memory. This delay depends on the size of the packet. Therefore at the output of the transfer model the time stamp of the beginning of the incoming packet is now $m'[n]=m[n]+D[n]$. In our approximation (lower section) the end of the packet is now used to indicate the time stamp of the packet, which is given by $m'[n] = m[n]+PacketSize/BitRate$. By using a delay of $PacketSize/BitRate$ to approximate $D[n]$ our model doesn't require knowledge of the actual packet transfer time, yet still maintains a packet size dependent delay.

\subsubsection{Interrupt Coalescence}
In a software based measurement system, after the packet data has been transferred to memory the NIC asserts a system interrupt informing the CPU that the packet is ready for processing. In low speed networks, an interrupt is asserted for each packet that the NIC receives, since at these speeds, e.g., 10 Mbps to 100 Mbps, the CPU is capable of responding to an interrupt service request (ISR) for each packet. However, in high speed networks e.g., 1 Gbps and beyond, the number of packets received increases significantly and as a consequence the number of interrupt service requests made by the NIC
increases as well. This can result in a condition called receive livelock where the host system cannot perform any processing and spends all of its clock cycles servicing the interrupt requests. To prevent livelock, a technique called interrupt coalescence is used in high speed NIC which reduces the
workload of the host CPU by lowering the number of interrupts issued by the NIC \cite{KSalah}.

\begin{figure}[ht]
	\centering
	\includegraphics[width = 5in]{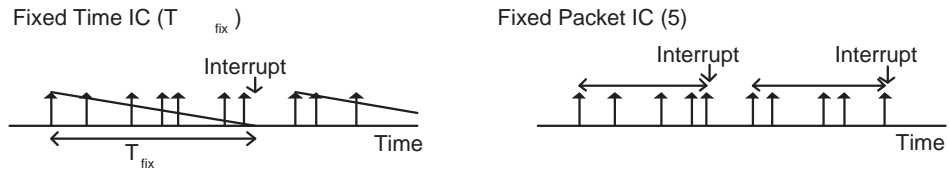}
	\caption{Example of interrupt coalescence using TIC and PIC methods}
	\label{fig:FixPackTime}
\end{figure}

Three variations of interrupt coalescence are commonly used: timer-based IC (TIC), packet-based IC (PIC), and hybrid time based methods (HIC), such as those found in the Intel NIC\cite{Intel, TimerBasedIM, KSalah}. Because the goal of interrupt coalescence is to reduce the interrupt rate imposed on the measurement system by the NIC, the obvious solution is to not issue an interrupt for each packet received. PIC does just this: after a packet arrives the NIC card waits until a given number of subsequent packets is received before an interrupt is asserted \cite{KSalah}. The limitation of this method is that during low traffic rate conditions the latency between the first packet arrival and interrupt assertion becomes excessive. Alternatively TIC waits a fixed amount of time after the arrival of a packet before issuing an interrupt. In low traffic rate situations long fixed time values also lead to excessive latency for TIC. Figure~\ref{fig:FixPackTime} shows how TIC and PIC generate interrupt service requests. On the left TIC is used with parameter $T_{fix}$, which is the length of time after the initial packet arrival before an interrupt is issued. Similarly on the right PIC is pictured where the number of packets received before asserting an interrupt is set to $5$.  

\begin{figure}[htb]
\centering
\includegraphics[width = 3in]{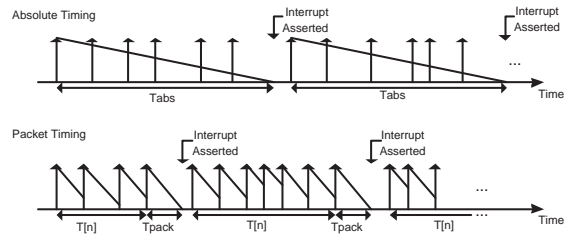}
\caption{Examples of interrupt coalescence with absolute and packet timers controlling the assertion of interrupts}
\label{fig:interrupt}
\end{figure}

Although these simple methods successfully reduce processor burden, their limitations create a need for HICs such as that developed by Intel, which incorporates two separate mechanisms to control the number of interrupts asserted on the system. These mechanisms are controlled by two separate timers, namely the absolute and packet timers\cite{Intel}. Figure \ref{fig:interrupt} illustrates the difference between the timer variations. The packet timer, $T_{pack}$, begins counting down after a packet has been received. If additional packets are received before the timer finishes counting down, then $T_{pack}$ is reset and begins counting down again. Anytime $T_{pack}$ reaches zero before a packet arrives to reset the timer, the NIC will assert an interrupt and all packets that were received since the previous interrupt will be serviced. During low traffic rate conditions the packet timer prevents excessive latency in the packet processing.

However, under high traffic rate situations when packets arrive in rapid succession it is possible that the packet timer will never count down to zero, so a second timer called the absolute timer, $T_{abs}$, is needed. Similar to fixed time IC, the absolute timer delays the assertion of the interrupt service request for a fixed interval following the receipt of an incoming packet. All packets received during the interval $T_{abs}$ are processed by the system during the same ISR as the initial packet. The absolute and packet timers work together to ensure interrupts are asserted efficiently during high and low traffic rate conditions.



\subsubsection{Software Induced Delay}
After interrupt coalescence generates an interrupt service request for a group of packets, each must be processed by the system. Processing, which includes time stamp generation, is performed individually for all packets in the group. Although the processing time is relatively short because it is done on individual packets the time stamps for the packets received during a single interrupt are generally not the same. The inter-arrival timing created between packets serviced by the same interrupt is artificial and does not reflect any of the characteristics of the original signal. Therefore it is up to the designer of the analysis task whether or not to treat the individual packet time stamps as being accurate or to remove the artificial inter-arrival timing in a way that does not remove important information about the signal. The artificial inter-arrival timing is essentially high frequency noise that is added to the signal. In order to reduce the effect of this high frequency noise we selected a more natural signal representation, which we describe in the following section.

\begin{figure}[htb]	
\centering
\includegraphics[height=1in]{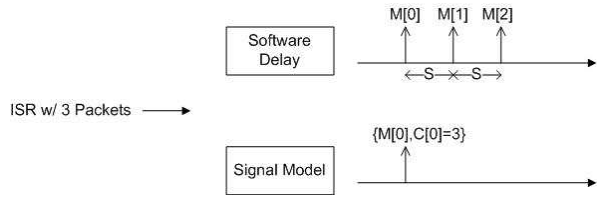}
	\label{fig:softmodel}
	\caption{Example of Software Induced Delay vs. Signal Model}
\end{figure}

Figure~\ref{fig:softmodel} shows how the delay in packet time stamping caused by the software creates artificial inter-arrival timings. In this example an ISR was asserted after IC, which contains 3 packets for processing. In the top portion of the figure a software delay, $S$ is introduced which is the processing time for the software to generate a time stamp for the individual packet. This creates an artificial inter-arrival time between measurements $M[0]$ and $M[1]$ as well as between $M[1]$ and $M[2]$ equal to $S$.  In the bottom portion of the figure our signal model, explained in the next section, treats packets which arrive during the same ISR as one measurement thus removing the effect of the artificial inter-arrival timing.

\section{Analysis}
\label{sec:analysis}
In this section we analyze the software measurement system further. First we consider what is the best way to represent the signal at the output of the measurement system. Because of timing inaccuracies created by interrupt coalescence and software delay, determining the best signal representation is important and not necessarily straight forward. Second we analyze in more depth a measurement system that uses hybrid interrupt coalescence. This IC method is selected because it is the most general of the systems described above. Specifically, HIC produces measurements where both the number of packets per measurement and the measurement inter-arrival times are variable. PIC and TIC therefore represent particular cases of HIC, where either the number of packets per measurement is fixed (PIC) or the measurement inter-arrival times are approximately fixed (TIC).
\subsection{Signal Representation}
\label{sec:signal}
The first step in improving analysis based on software measurements is to redefine the signal representation. A standard uniformly sampled signal representation is not appropriate for software based measurements because, as mentioned previously, measurements are triggered by the signal being measured, i.e. the measurement timings are signal dependent. Due to interrupt coalescence, multiple packets are delivered to the system jointly and artificial inter-arrival timing is created by the software that generates the packet time stamps. This artificial inter-arrival timing is unrelated to the actual packet timing and is a source of error. Thus it is desirable to remove software induced delay and to do so the most natural to represent the signal as a vector:
\begin{equation}
 X(k) = \{M(k), C(k)\}
 \end{equation}
where $M(k)$ is the time stamp of the $k$th group of packets measured and $C(k)$ gives the number of packets in the group. This represents precisely the measured information: each interrupt provides timing information and a packet count. 

This redefined signal representation models the data at the last stage of the measurement system. The entire measurement system consists of the three modules described above: packet transfer, interrupt coalescence, and software induced delay. Figure~\ref{fig:measurementsys3} shows how a collection of packets traverses the measurement system, and how the final output measurement is achieved. The packet transfer model was created to approximate the time required transfer a packet to memory. Our model uses a delay equal to the size of the packet in bits divided by the bit rate of the incoming link. This allows us to model the packet size dependence of the delay without actual knowledge of the packet transfer time. Next we subject this delayed signal to interrupt coalescence, which as mentioned previously is a timing dependant delay which generates an ISR signaling to the host system that it can begin processing the packets. In order to avoid the artificial inter-arrival timing created in software, in the software induced delay model, we use the vector signal representation, $X[k]=\left\{M[k],C[k]\right\}$ where $M[k]$ represents the time stamp of the first packet generated by the system, and $C[k]$ gives the number of packets received simultaneously from the corresponding ISR. The selected signal representation removes unwanted noise from the signal while, with proper processing, retains much of desired signal properties for analysis.


\subsection{Further Analysis of HIC}
\label{sec:HIC}
With the new signal representation that was derived in the previous section we proceed to analyze specifically hybrid interrupt coalescence. We focus on HIC because the analysis is more general than could be done for PIC or TIC, as these can be seen as particular cases of HIC where either the inter-arrival times or the number of packets are fixed. 

\subsubsection{Ideal Poisson Process}
\label{sec:IPP}
In order to understand the effect of HIC we consider an idealized case where the input to the measurement system is a signal with exponentially distributed inter-arrival times (with scale parameter $\lambda$), and the output is a new signal with our vector signal representation. Similar to \cite{ToCoalesce}, we start with an ideal input to gather intuition about the measurement system and develop analytic results for the measurement system output. Although what is presented here is specific to an ideal Poisson input process we feel that the analysis presented illustrates the biases that one would encounter when trying to estimate the input parameters of actual Internet traffic using the measurement system output. 

In the following analysis we use the variable $Y$ to indicate the inter-arrival time between measurements, and use a subscript on the variable to indicate either: the number of packets contained in the measurement or the distribution the inter-arrival time is derived from. Also we indicate estimates of random variables by the $\hat{}$ operator, e.g. $\hat{\lambda}$ is an estimate of the actual scale parameter $\lambda$. Further, we make use of two variations of the exponential distribution, specifically the shifted exponential distribution and the right-side truncated exponential distribution. As defined in \cite{olive} the probability density function (PDF) and expected value for the shifted and right-side truncated exponential distributions are given by:
\begin{equation}
\begin{gathered}
	f_{Y_{Shift}}(y|\lambda,b) = \frac{e^{-\left(\frac{y-T_{pack}}{\lambda}\right)}}{\lambda} \\
	E[Y_{Shift}] = \lambda_{Shift} = \lambda + b	
	\label{eq:shiftedexpdist}
	\end{gathered}
\end{equation}
where $\lambda$ corresponds to the scale parameter of the input sequence and the support for the shifted exponential distribution ranges from $[b, \infty]$.

\begin{equation}
\begin{gathered}
	f_{Y_{Trunc}}(y|\lambda,b) = \frac{e^{-\frac{y}{\lambda}}}{\lambda (1-e^{-k})} \\
	E[Y_{Trunc}] = \lambda_{Trunc} = \lambda \left[\frac{1-(k+1)e^{-k}}{1-e^{-k}}\right]
	\label{eq:truncatedexpdist}
	\end{gathered}
\end{equation}
again $\lambda$ is the input scale parameter, $k = \frac{b}{\lambda}$ and the support is defined on $[0,b]$.   

\begin{figure}
  \centering
  \subfloat[One Packet]{\label{fig:onepack}%
    \includegraphics[width=0.45\textwidth]{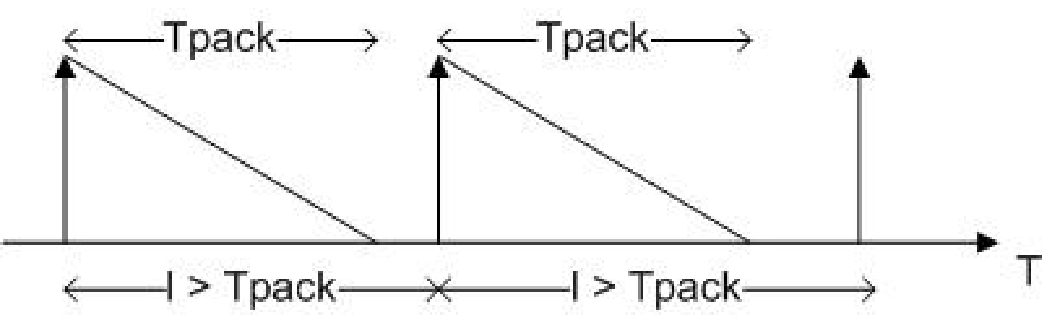}}%
  \quad%
  \subfloat[Two Packet]{\label{fig:twopack}%
    \includegraphics[width=0.45\textwidth]{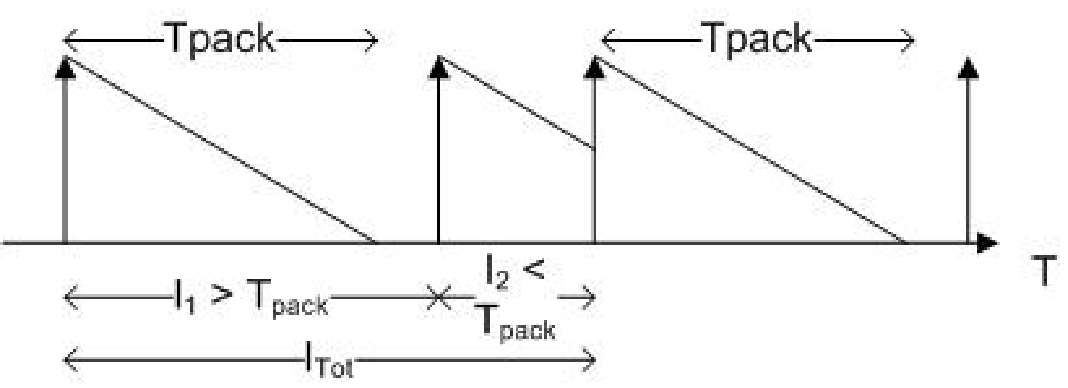}}
   \caption{Examples featuring one packet per measurement~\ref{fig:onepack} and two packets per measurement~\ref{fig:twopack}}
  \label{fig:multiex1}
\end{figure}

Consider first two consecutive measurements each of which contains only a single packet. Because each measurement only contains one packet each ISR that triggered the measurement must have been generated by $T_{pack}$. The time-stamp for each measurement is then the original time that the packet arrived at the NIC plus some fixed delay ($T_{Pack}$ plus system processing delay), which we assume is constant. The inter-arrival time between the measurements therefore corresponds to the actual inter-arrival time of the original packets. If we consider only such packet pairs and use them to estimate the expected value of the inter-arrival times, it would be reasonable to guess that expected value would be an accurate estimate of the mean of the input inter-arrival times, i.e. 
\begin{equation}
\begin{gathered}
\hat{\lambda} = \frac{1}{N}\sum_{\forall n:C[n]=C[n-1] = 1}{[M[n]-M[n-1]]}  \\
N = \sum_{\forall n:C[n]=C[n-1] = 1}{1}\\
\hat{\lambda} \approx \lambda
\end{gathered}
\end{equation}
However, due to the way the packet timer works it is only possible to get back-to-back single packet measurements if the inter-arrival time is strictly greater than $T_{pack}$, as shown in Figure~\ref{fig:onepack}. Thus the inter-arrival times we consider are all strictly greater than $T_{pack}$ and are not exponentially distributed but instead belong to the shifted exponential distribution. Clearly our estimate of the input scale parameter is wrong, but by knowing the value of $T_{pack}$ we can correct our estimate by using the expression for expected value above. Letting $b=T_{pack}$ in Equation~\ref{eq:shiftedexpdist} we find that $\hat{\lambda}_{shift} =\hat{\lambda} + T_{pack}$. Therefore by subtracting $T_{pack}$ we obtain an estimate of the input scale parameter, $\hat{\lambda} = \hat{\lambda}_{Shift} - T_{Pack}$.

Next we consider a slightly more complex situation, measurements that contain two packets. Specifically, $Y_2 = \{M[n] - M[n-1]:C[n] = 2\}$. The inter-arrival time of a measurement with two packets is the sum of two inter-arrival times as depicted in Figure~\ref{fig:twopack}.  The first inter-arrival time is between the last packet in the previous measurement and the first packet in the current measurement (indicated by $I_1$ in Figure~\ref{fig:twopack}). This inter-arrival time must be greater than $T_{pack}$, otherwise the subsequent packet would be coalesced with the previous packet. Therefore this inter-arrival time falls into the shifted exponential distribution case. The second inter-arrival time is between the two packets in the measurement ($I_2$ in Figure~\ref{fig:twopack}), which must be strictly less than $T_{pack}$ otherwise the packet timer would have generated an interrupt. This second type of inter-arrival time is not exponentially distributed but belongs to the right-sided truncated exponential distribution. 

The inter-arrival time of measurements containing two packets, i.e. $Y_2$, is then the summation of two random variables, $Y_2 = Y_{shift} + Y_{trunc}$. Therefore when estimating the expected value of inter-arrival times for measurements containing two packets it is expected that $\hat{\lambda}_2 = \hat{\lambda}_{Trunc} + \hat{\lambda}_{Shift}$. However based on simulation this is actually not the case. Instead the expected value of the inter-arrival times of measurements with two packets is: $\hat{\lambda}_2 = \hat{\lambda}_{Shift} + \hat{\lambda}$. This is likely caused by the ratio of $\lambda$ and $T_{pack}$. When $k$ is large, which occurs when the value of $T_{Pack}$ is too large for the input traffic rate, then $\lambda_{Trunc}\cong\lambda$.

To complete our analysis we consider the distribution of inter-arrival times for \emph{all} measurements (i.e. any number of packets per measurement). To do this we first find the PDF of the inter-arrival times of the measurements with two packets. This is found to be:
\begin{equation}
\begin{gathered}
	Y_2 = Y + Y_{Shift}\\
	f_{Y_2}(y_2|\lambda, T_{pack}) = f_Y(y|\lambda)\ast f_{Y_{Shift}}(y_{shift}|\lambda, T_{pack})\\
	f_{Y_2}(y_2|\lambda,T_{pack}) = \frac{(y_2-T_{pack})e^{-\left(\frac{y_2-T_{pack}}{\lambda}\right)}}{\lambda^2}
\end{gathered}	
\end{equation} 
Notice that we consider the sum of one inter-arrival time from the shifted exponential distribution, and one from the standard exponential distribution instead of the truncated exponential distribution. This is done to be consistent with the observation made above. 

Similarly we can compute the PDF of the inter-arrival times of the measurements with $i$ packets as:
\begin{equation}
\begin{gathered}
	Y_i = Y + Y_{i-1}\\
	f_{Y_i}(y_i|\lambda, T_{pack}) = f_Y(y|\lambda)\ast f_{Y_{i-1}}(y_{i-1}|\lambda, T_{pack})\\
	f_{Y_i}(y_i|\lambda, T_{pack}) = \frac{(y_i-T_{pack})^{i-1}e^{-\left(\frac{y_i-T_{pack}}{\lambda}\right)}}{\lambda^i(i-1)!}
\end{gathered}
\end{equation} 

When considering the distribution of inter-arrival times for all measurements, we have the following probability distribution:
\begin{equation}
P(y = \hat{y}) = P_{Y_{Shift}}(y_shift=\hat{y}) + \sum_{i=2}^N{P_{Y_i}(y_i=\hat{y})}
\end{equation}
where N is some maximum value for the number of packets in a measurement (this is arbitrary). Therefore:
\begin{equation}
\begin{gathered}
	P(y = \hat{y}) = \frac{e^{-\left(\frac{\hat{y}-T_{pack}}{\lambda}\right)}}{\lambda}+\sum_{i=2}^N
	{\frac{(\hat{y}-T_{pack})^{i-1}e^{-\left(\frac{\hat{y}-T_{pack}}{\lambda}\right)}}{\lambda^i(i-1)!}}\\
	P(y = \hat{y}) = \frac{e^{-\left(\frac{\hat{y}-T_{pack}}{\lambda}\right)}}{\lambda}\sum_{i=0}^N
	{\frac{\left(\frac{\hat{y}-T_{pack}}{\lambda}\right)^{i}}{(i)!}}\\	
	N\rightarrow \infty\\
	P(y = \hat{y}) = \frac{e^{-\left(\frac{\hat{y}-T_{pack}}{\lambda}\right)}e^{\left(\frac{\hat{y}-T_{pack}}{\lambda}\right)}}{\lambda}\\
	P(y = \hat{y}) = \frac{1}{\lambda}	
\end{gathered}
\end{equation}

Thus if we consider inter-arrivals of measurements with any number of packets the distribution converges to a uniform value. This becomes important when we develop our detection method where we use the uniformly distributed assumption to distinguish ``random'' background traffic from background traffic combined with periodic attack traffic. Note that convergence above also applies when $Y_2 = Y_{shift} + Y_{trunc}$ instead of the approximation made in the above derivation, i.e., $Y_2 = Y + Y_{trunc}$. Further the above convergence also applies to an ideal Poisson process. Taking higher-order differences of exponentially distributed inter-arrivals times produces inter-arrivals that are n-Erlang distributed. Considering inter-arrivals with any number of packets (i.e., summing the n-Erlang distributions over $n\in[1,\infty]$) the distribution also converges to $\frac{1}{\lambda}$.

Although our analysis considered an ideal Poisson input sequence some of the results are more general. The first result shows that estimates of input traffic parameters calculated using the output of the measurement system are going to be biased. Further the bias is different depending on if the measurement contains one packet versus measurements with multiple packets. This makes intuitive sense because the measurements containing a single packet are precise, the inter-arrival time is exact, yet they are left truncated. Conversely the measurements with multiple packets contain a single precise inter-arrival time combined with multiple imprecise right-truncated inter-arrivals. Therefore when using the output of the measurement system for estimating parameters of the input sequence knowledge of the potential inaccuracies should be understood in advance.

We can make one final observation regarding parameter estimation using the output of the measurement system. In order to estimate the input rate parameter $\lambda$ it suffices to treat the output of the measurement system as an n-Erlang distribution where n is itself random. Letting the output inter-arrival measurements be represented by the random variable $M = erlang(N,\lambda)$ then:
\begin{equation}
E[M] = E_C[E_{\lambda}[M |C=c]] = E_C[C\cdot\lambda] = \lambda \cdot E_C[C]	
\end{equation}
Thus the expected value of the output inter-arrival sequence, without regard to number of packets per measurement C, is exactly $\lambda$ times the average number of packets per measurement at the output. Consequently we can estimate the input rate parameter by
\begin{equation}
	\hat{\lambda} = \frac{\frac{1}{K}\sum_{i=1}^K{M[i]}}{\frac{1}{K}\sum_{i=1}^K{C[i]}} 	
\end{equation}
without concern for shifted or right-truncated exponential distributions.

\section{Measurement System Task}
\label{sec:task}
The detection task we are trying to design a method for is to detect periodic packets in aggregate traffic. The periodic packets could be created by a denial-of-service (DOS) attack or correspond to the existence of a bottleneck link \cite{xinming}. The attack packets are spaced periodically, with period $P$ seconds, because of two possible scenarios. The first is that the code generating the attack packets spaces the packets at equal intervals. This can be the result of the algorithm specifically outputting packets at a fixed period, or the algorithm attempting to output packets as fast as possible while the host operating system (OS) or central processing unit (CPU) limits the packet output rate and outputs packets periodically \cite{Hussain03c}. In the second scenario the attack packets encounter a bottleneck link upstream. The bottleneck link outputs packets back-to-back at its maximum rate. Therefore the time between attack packets is constant and $P$ is equal to the time it takes to output one attack packet at the bottleneck link speed. If the attack packets flow into a network with a higher link speed, then the time between attack packets is still equal to $P$ even though the attack packets are no longer back-to-back. 

The above two scenarios have been considered previously. In \cite{Hussain03c} the authors consider spectral characteristics of DOS attacks, which is shaped by the program used to generate the attack or the host system OS and CPU. Then in \cite{Hussain06a} the spectral characteristics are used to 'fingerprint' the attacks and identify repeated attacks bearing the same fingerprint. The second scenario mentioned above is the basis for packet pair techniques that measure the end-to-end available bandwidth of an internet link. In \cite{Prasad03bandwidthestimation} the authors explain how by evaluating the inter-arrival time between pairs of probe packets sent out over an internet link it is possible to measure the available bandwidth on that link. The work in \cite{xinming} applies the same principles found in packet pair techniques, but instead uses spectral techniques to identify bottleneck links without the use of packet probes. 

Although we assume that the attack packets are at one point periodic the task is to detect them in aggregate traffic. This involves merging the attack traffic with background internet traffic, which does not preserve the exact periodic structure of the attack packets. We can however assume that the time between attack packets is approximately equal to $P$ as any variation due to queuing with other packets are likely to be small. Further we can assume that in a fixed time interval the average number of attack packets is going to be constant.

\subsection{Measurement System Configuration and Test Data}
\label{sec:Config}
In the analysis to follow we use two distinct measurement system configurations to highlight the performance difference between our detection method, Periodic Detection using Multiple Measurements (PDMM), and another capable periodic detection system, Periodic Attack Detector (PAD). The two configurations were selected specifically to produce measurements with desired characteristics. The first system, HICv1, was selected to produce measurements with significant variability in the number of packets per measurement, i.e. $C[n]$, as well as in the inter-arrival time between measurements, $M[n]-M[n-1]$. HICv2 on the other hand was configured such that the majority of measurements would be generated by the absolute timer, and thus the variability in the inter-arrival times would be much less (closer to uniformly sampled signal). The system parameters for HICv1 are $T_{abs} = 300 \:\mu$Seconds and $T_{pack} = 30 \:\mu$Seconds while the parameters for HICv2 are $T_{abs} = 120\: \mu$Seconds and $T_{pack} = 33\: \mu$Seconds. Note that the measurement system settings were specifically selected to produce measurements with the desired characteristics based on prior knowledge of the input traffic rate. The identical system settings are unlikely to produce the same measurement characteristics for different input traffic rates, which is clearly shown in Section~\ref{sec:experiment} when the above system parameters produce different measurement characteristics for the high and low rate traces.  

The test data used in the analysis of the detection methods consists of synthetic traces\footnotemark which we created by combining real background internet traffic at two different rates, 320 and 196 Mbps, with simulated attack traffic at a rate of either 30 Mbps or 15 Mbps, using the stream merger application \cite{Kamath02a}. Specifically the high rate background traffic was combined with the higher 30 Mbps rate attack while the lower rate background traffic was merged with the lower rate (15 Mbps) attack. For the software based measurement systems the effect of interrupt coalescence was simulated on the above trace data before the data was formatted for the proper signal representation. 

The average packet rate for the combined input traces was around $80,000$ packets per second for the high rate trace (320 Mbps background + 30 Mbps attack) and around $47,000$ packets per second for the low rate trace (196 Mbps background + 15 Mbps attack). At the output of the measurement system the average measurement rate was approximately $11,000$ measurements per second for both measurement systems and for both input traces. Clearly the effect of interrupt coalescence is more apparent on the high rate trace. 
\footnotetext{All traces are provided by the LANDER project and available through PREDICT (https://www.predict.org) \cite{syntheticTrace}}

\subsection{Periodic Attack Detector (PAD)}
\label{sec:PAD}
\begin{figure}[htb]
	\centering
  \includegraphics[width=0.75\textwidth]{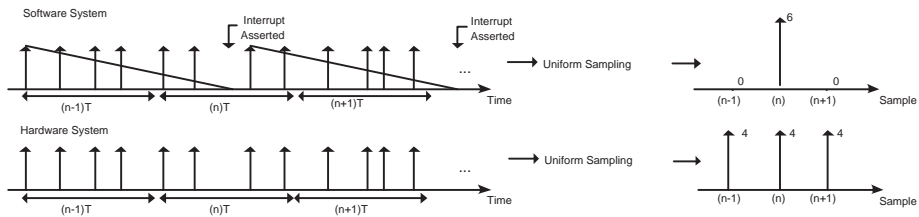}
  \caption{Example of signal distortion due to interrupt coalescence}
  \label{fig:sigdist}
\end{figure}

Before discussing our method, which we design specifically for the measurement system, we motivate the need for such design by showing how the measurement system can impact the performance of analysis techniques which assume a typical uniformly sampled signal. We focus on a particular detection method called the PAD that was presented in \cite{xinming, Thatte08a}. The PAD was chosen because it uses standard signal processing techniques, such as computation of the power spectral density (PSD). Because the PAD computes the PSD of the signal the performance of the detection scheme depends on the precision of the input signal. In a hardware based measurement system this is not an issue as the precision of typical DAG measurement cards ranges from $10 \mu$Sec to $.1 \mu$Sec \cite{highquality}. However, when the measurement system creates errors in the signal timing standard signal processing tools like the PSD are not as accurate and can degrade performance of detection methods that rely on them. Given the inaccuracies in the measurements if a uniformly sampled signal is generated in the same manner as for the hardware based system the resulting signal will be distorted. An example of a potential distortion is shown in Figure \ref{fig:sigdist} where the uniformly sampled representation of the signal is different between the software (top figure) and hardware systems (bottom figure) because of interrupt coalescence. 

To show the degradation in performance of PAD we evaluated the time to detection for PAD using a signal generated from hardware and software system measurements. The software signal was generated as described in Section~\ref{sec:Config}, and the hardware signal was generated using the same synthetic trace without applying the simulated software measurement system. The time to detection results for PAD using both the high rate input traffic (320 Mbps background + 30 Mbps attack) and the low rate input (196 Mbps background + 15 Mbps attack) are shown in Table~\ref{Tab:PAD}.
\begin{table*}[htbp]
	\centering
		\begin{tabular}{|l|c|c|c|}
		\hline 
		Measurement System&Hardware&HICv1&HICv2\\
		\hline
		High Rate Time to Detection (Seconds) & .627 & 1.731 & 1.110 \\ \hline
		Low Rate Time to Detection (Seconds) & .803 & - & - \\ \hline
		\end{tabular}
		\caption{PAD detection performance comparing hardware and software measurement systems}	
		\label{Tab:PAD}
\end{table*}
Clearly there is significant degradation in the performance of the PAD when using the software based measurements. Especially in the low rate traffic scenario where PAD was unable to detect an attack in the 20 second window we allowed in the experiment. The likely cause of the degradation is that the characteristics of the measurement system were not considered in the design of PAD. Specifically, PAD assumes an accurate, uniformly sampled signal as input which is not the case following the software measurement system. In the high rate input experiment PAD performed best when the measurement system settings dictated that $T_{abs}$ generate most of the interrupts. When $T_{abs}$ generates most of the interrupt the timing between measurements becomes more uniform, and PAD performs better. However, when the input traffic is low rate then neither of the two measurement system settings cause $T_{abs}$ to generate most of the interrupts. This causes the timing between measurements to be highly randomized because they are controlled by $T_{pack}$ asserting interrupts, and interrupts asserted by $T_{pack}$ are signal dependent. 


\section{Detection Using Multiple Measurements and the Pearson Chi-Square Test}
\label{sec:chi}
From the results in the previous section it is clear that the measurement system affects the performance of standard detection methods. Our goal in this section is to design a detection method using our knowledge about the measurement system. 
\subsection{Developing Intuition}
\begin{figure}[htb]
	\centering
  \includegraphics[width=0.45\textwidth]{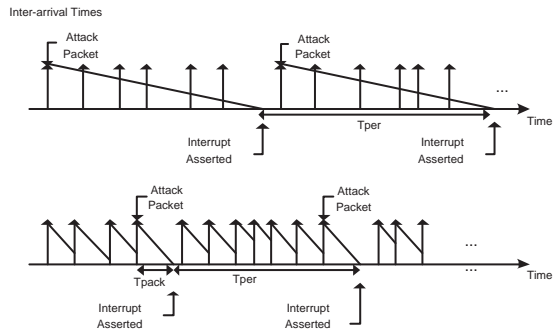}
  \caption{Two examples of inter-arrivals equal to attack period due to interrupt coalescence}
  \label{fig:interarrival}
\end{figure}

The intuition behind our detection method begins with a pair of observations. First we noticed that in aggregate traffic containing an attack inter-arrivals times between consecutive groups of packets were equal to the attack period with probability greater than chance, where inter-arrival times are calculated as $I[n] = M[n] - M[n-1]$. This occurs because the ISRs that generate the time stamps are controlled by the packets, i.e., are signal dependent. By considering how interrupt coalescence works we see that multiple situations exist where inter-arrival times would equal the attack period, as illustrated by the two examples in Figure~\ref{fig:interarrival}. 

In the first example the initial packet in each group is an attack packet, and the interrupt for each group is asserted by the absolute timer. Assuming the period of the attack is longer than the absolute timer, $P > Tabs$, then the inter-arrival time between the consecutive groups of packets will be equal to the period, $P$. In the second example the final packet in each group is a periodic packet and assuming the packet timer expires after the periodic packet in both groups then the inter-arrival time of the groups of packets will be equal to the period of the attack signal. Further by considering more than just back-to-back inter-arrivals (to be discussed below) many more scenarios similar to these two simple examples occur, which create inter-arrival times equal to the attack period, $P$.

The second observation we made regards the background traffic. We found that if we considered not only the first order inter-arrivals but higher order inter-arrivals as well, then the histogram of inter-arrivals is approximately uniform over a fixed range. Higher degree inter-arrivals refers to taking the difference in time stamps of measurements further apart than consecutive, i.e. $I_i(n) = M(n) - M(n - i)$ where $i > 1$ is the order of the difference. The intuition behind why the distribution of background traffic inter-arrivals is approximately uniform can be shown from the analysis in section~\ref{sec:IPP}. Paxson \cite{Paxson95a} and many others have shown that background internet traffic is not an ideal Poisson process. However, based on significant analysis of background traffic we found that the histogram of inter-arrivals is sufficiently uniform to not generate false positives for our detection method described below. 

\subsection{Detection Method}
Based on the above observations we designed a detection method that analyzes the distribution of first and higher order inter-arrival times between groups of network packets. The first step in out method is to generate an inter-arrival sequence, $I_i(n)$ for $i = 1, 2,..., N$ from the measurements time stamps $M[n]$. Currently our method processes data in blocks of length $L$, and uses the first block to populate the distribution before detection is performed. Thus for the first $L$ measurements an inter-arrival sequence is constructed. Then a histogram for the values in the inter-arrival sequence is populated. 

Once the histogram is initially populated the data is processed a block at a time using the same procedure: generate inter-arrival sequence for block of data, and use the inter-arrival sequence to populate the histogram. Starting with the second block of data, detection is performed after the new data is added to the histogram. Based of the second observation above we use a hypothesis test where the null hypothesis is that the distribution is uniform. The Pearson Chi-Square test, a common hypothesis test used to check goodness of fit to a known distribution, is used to evaluate the histogram data to see how well the inter-arrivals fit a uniform distribution.  

The Pearson Chi-Square test first segments the histogram data into $K$ sub-bins. Then using the uniform distribution assumption the expected number of occurrences of inter-arrivals in each sub-bin is $E = O/K$ where $O$ is the sum of all the points in the histogram. Next a chi-square distributed random variable is computed using:
\begin{equation}
	\chi^2 = \sum_{k=1}^K{\frac{(O_{k} - E)^2}{E}} 	
\end{equation}
Where $O_{k}$ is the actual number of occurrences in the $kth$ sub-bin. 

Finally the Pearson Statistic, $p$, is calculated, which is essentially the value of the CDF of a Chi-Square distribution with $K-1$ degrees of freedom computed at the value $\chi^2$. The null hypothesis, that the data does fit the uniform distribution, is rejected if $1-p < T $ where $T$ is used to control the probability of false alarm and is typically selected to be 0.05. 

Pseudocode of our detection method is given below. The detection first populates the histogram using a block of measurements. After the histogram is populated the method process the incoming measurements in blocks of $L$ measurements. Processing involves updating the histogram, and then computing the value of the Chi-Square random variable using Pearson's method. Finally, the value of the Pearson statistic, $p$, is evaluated and compared against the user selected threshold $1-T$. If an attack is detected the algorithm halts. If no attack is detected the algorithm proceeds to process the next block of $L$ measurements. 

\begin{verbatim}
\*Initialize Histogram*\
for i = 1:N \*N=Order of Inter-arrivals*\
  for n = i:L \*Initialize Histogram for L measurements*\
    CurDif = M[n] - M[n-i]
    Histogram(CurDif)++
  end    
end
\*Begin Detection*\
while(1)
  for i = 1:N \*Order of Inter-arrivals*\
    for n = i:L \*Update Histogram for L measurements*\
      CurDif = M[n] - M[n-i] 
      Histogram(CurDif)++
    end
  end
  NumBins = K
  O = sum(Histogram)
  E = O/NumBins
  BinWidth = length(Histogram)/NumBins
  ChiSq = 0
  \*Compute value of ChiSq Random Variable*\
  for j = 1:NumBins
    O(j) = sum(Histogram((j-1)*BinWidth:j*BinWidth))
    ChiSq = ChiSq + (O(j)-E)^2/E
  end
  \*Compute Pearson Statistic using ChiSq pdf with K-1
    Degrees of Freedom*\
  p = chisqpdf(ChiSq, NumBins - 1)
  if (p > 1-T)
    \*Detected Attack*\
    break;
  end	
end
	
\end{verbatim}

There is one caveat to our method which is that due to system effects, inter-arrivals below $T_{abs}$ are not uniformly distributed. Specifically many inter-arrivals occur near $T_{pack}$ and $T_{abs}$. Because of this the detection method ignores inter-arrivals shorter than $T_{abs}$. However, this does not impact our ability to detect periodic signals with $P<T_{abs}$. If we take inter-arrivals of sufficient order then our method should still be able to detect a harmonic multiple of the base period that is larger than $T_{abs}$, i.e. instead of detecting $P$ we can detect $kP > T_{abs}$. There also is an upper limit on the length of the inter-arrivals considered, which is selected such that we can still capture the attack period of very low rate attacks. The upper limit, $T_{Max}$ must be selected such that $T_{Max} > P$. Selecting a larger value of $T_{Max}$ often requires taking a higher order differences in inter-arrival times as well, because for larger $T_{Max}$ the higher order inter-arrival differences are required to produce a near uniform distribution of the histogram.

\subsection{Generic Detection Example}
Here we examine detection of a single periodic peak without harmonics in background traffic with a perfect uniform distribution of inter-arrival times. We consider the percent deviation, $S$, in a particular sub-bin required to detect the periodic peak. Let $K$ be the total number of sub-bins, and $O$ be the total number of points in the histogram. Then the expected number inter-arrivals in each sub-bin is:
\begin{equation}
	E = \frac{O}{K} 	
\end{equation}

The actual number of occurrences in the sub-bin with the periodic peak is:
\begin{equation}
	O_{i} = (\frac{1}{K}+S)\cdot O 	
\end{equation}
and the actual number of occurrences in the remaining bins is:
\begin{equation}
	O_{j} = (\frac{1}{K}- \frac{S}{K-1})\cdot O \ \ j\neq i	
\end{equation}

Computing the chi-square random variable:
\begin{equation}
\begin{gathered}
	\chi^{2} = \frac{1}{E}\left[(O_i - E)^2 + \sum_{j\neq i}{(O_{j} - E)^2}	\right]\\
	\chi^{2} = \frac{K}{O}\left[(S\cdot O)^2 + (K-1)\cdot(\frac{S\cdot O}{K-1})^2	\right]	\\
	\chi^{2} = S^2\cdot O (K+\frac{K}{K-1})
\end {gathered}
\end{equation}

What we see from the above equation is that the chi-square value depends on: the total number of measurements and the percent deviation from uniformity created by the periodic peak. The value of $\chi^2$ required to exceed the null hypothesis threshold, $T$, depends on the number of degrees of freedom of $\chi^2$ which is $K-1$, so increasing the number of sub-bins is not an efficient strategy for increasing detection performance. Notice that the value of $\chi^2$ depends on $S^2$ so larger deviations will be detected quicker geometrically instead of linearly. 

\section{Experiments}
\label{sec:experiment}
To examine the performance of our detection method we compare it to the performance of the PAD detection method for a few input traffic and measurement system combinations. The input traffic and measurement system combinations were specifically selected to highlight how certain traffic characteristics affected the performance of the detection methods. To show the traffic characteristics quantitatively we consider the variance in the inter-arrival times (first order only) of the measurements at the output of the measurement system. 

The input traffic signals and measurement system configurations used in the experiments were described in section~\ref{sec:Config}. First we consider the high rate signal (320 Mbsp background + 30 Mbps attack). Using the high rate signal as the input to our two measurement systems the signal characteristics at the output were very different. Following the HICv1 system the variance in the measurement inter-arrival times was $2.78$ Nano-Seconds$^2$, and for the HICv2 system the variance was much smaller at $1.37$ Nano-Seconds$^2$. The ratio between these values is approximately $2$. Note that the mean in the inter-arrival times between measurements is nearly identical for both output signals at around $90$ Micro-Seconds. The larger variance in the measurement inter-arrivals for HICv1 indicates that the inter-arrival times are more random. This should benefit PDMM which uses the information in the inter-arrival times to perform detection. The lower variance value found in the measurements from HICv2 indicates that the inter-arrivals are less random, and more like the uniform sample intervals that PAD expect. 

Because our detection method is specifically designed for the software measurement system output we do not test it using the hardware system signal.
The time to detection results for the high rate signal using PAD and PDMM detection methods are shown in Table~\ref{Tab:High}.

\begin{table*}[htbp]
	\centering
		\begin{tabular}{|l|c|c|c|}
		\hline 
		Measurement System&Hardware&HICv1&HICv2\\
		\hline
		PAD - Time to Detection (Seconds) & .627 & 1.731 & 1.110 \\ \hline
		PDMM - Time to Detection (Seconds) & - & .62 & 1.31 \\ \hline
		\end{tabular}
		\caption{Detection performance for PAD and PDMM using the 320 Mbps background + 30 Mbps attack signal}	
		\label{Tab:High}
\end{table*}

There are a few interesting details in the results of the first experiment. The first is that given the software system output that was designed to be more 'random' (HICv1), having a higher inter-arrival time variance, our method detects the attack much quicker than PAD and even rivals the performance of PAD using hardware measurements. Thus we achieve our goal of designing a detection method that by considering the measurement system can outperform a generic detection method. Second, given the measurement system configuration designed to produce measurements that are more like a uniformly sampled signal, thus having a lower inter-arrival time variance, PAD outperforms our method. This second point shows how important it is to consider the effect of the measurement system. By altering the measurement system configuration it is possible to operate in regions where the output signal characteristics benefit one detection method over the other. Our detection method was designed for when the measurement system was operation under normal conditions, where the interrupts were generated primarily by $T_{pack}$, yet in more extreme conditions, interrupts generated mainly by $T_{abs}$, the performance of our method degrades and PAD improves. 

Next we consider detection performance for the two mechanisms when using the low rate signal (196 Mbps background + 15 Mbps attack) as input to the measurement system. After the HICv1 system the variance in the measurement inter-arrival time was much smaller at $2.08$ Nano-Seconds$^2$, while at the output of the HICv2 system the variance was similar to that for v1 at $1.78$ Nano-Seconds$^2$. For the low rate input signal the ratio of the variances is much closer to unity at $1.16$. The cause of the change in characteristics from the high rate case are due to the fact that the input traffic rate is no longer high enough to force HICv2 into the extreme condition where most interrupts are generated by $T_{abs}$. 

Based on characteristics of the output of the measurement system we could expect that the detection performance of PDMM would be significantly better than PAD. This is indeed the case as is shown in Table~\ref{Tab:Low}. Notice that for both measurement systems while using the low rate input PAD was unable to detect an attack over the 20 Second window of time that we allowed for detection.

\begin{table*}[htbp]
	\centering
		\begin{tabular}{|l|c|c|c|}
		\hline 
		Measurement System&Hardware&HICv1&HICv2\\
		\hline
		PAD - Time to Detection (Seconds) & .803 & - & - \\ \hline
		PDMM - Time to Detection (Seconds) & - &.97&.88\\ \hline
		\end{tabular}
		\caption{Detection performance for PAD and PDMM using the 196 Mbps background + 15 Mbps attack signal}	
		\label{Tab:Low}
\end{table*}

The second experiment also has a few interesting conclusions. The first is that the results strengthen our previous argument that one must consider the effect of the measurement system. In this experiment the measurement system produced an output signal that was better suited for PDMM, and the detection performance showed this. Further, and different from the first experiment, the output signal characteristics at the output of both measurement systems were nearly identical and the time to detection for PDMM was nearly identical for the two output signals. The second conclusion that can be drawn from this experiment is that if one desires to select the measurement system configuration for a given detection method then the input signal must be taken into consideration. Looking at HICv2, in the first experiment the input traffic rate was high enough to force the measurement system into an extreme condition where most interrupts were generated by $T_{abs}$. In this situation PAD was the better choice for detection. However, in the second experiment with the same measurement system configuration the input traffic rate was not high enough to force the extreme condition and PAD was a poor choice for detection. 

Based on the results of the above experiments it is clear that the measurement system has a significant affect on the detection performance for both mechanisms. Additionally, the relationship between the input traffic and measurement system configuration must also be considered as this interaction determines the relevant characteristics of the output signal which effect detection performance.

\section{Conclusion}
\label{sec:conclusion}
We began this paper by examining the operation of typical software measurement systems. It was seen that measurement systems that employed timer-based or packet-based interrupt coalescence were less desirable than the hybrid method of IC due to latency issues under certain input traffic situations. Because of this we focused our attention on the hybrid method of interrupt coalescence and developed a deeper understanding of its effect on the timing of the input traffic signal. This led us to derive a signal representation at the output of the software measurement system that better conveyed the important signal characteristics. With our increased understanding of the measurement system and improved signal representation we next examined a typical analysis task, periodic signal detection. 

While analyzing a generic periodic signal detection method, PAD, we found that the detection performance degraded significantly between detection using a hardware system measured signal and one measured using a software based system. Using our knowledge of the hybrid measurement system we derived a new detection method, PDMM, specifically for the signal representation at the output of the measurement system. Through a select set of experiments we learned a few important lessons about the relationship between input traffic, measurement system configuration, and detection method. 

\begin{figure}[htb]
	\centering
  \includegraphics[height = 1.6in]{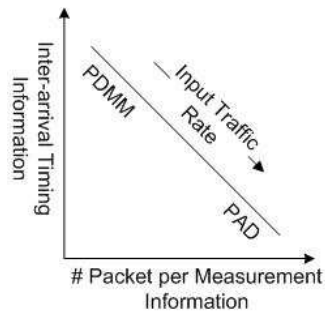}
  \caption{Graph showing where relevant measurement information is contained as input traffic rate varies}
  \label{fig:graph}
\end{figure}

The first lesson learned was that with careful consideration of the measurement system we can significantly improve detection performance compared to a generic detection method. Second, we learned that under certain input traffic and measurement system configurations the output of the measurement system will obtain characteristics that are amenable to detection using PAD. As mentioned in Section~\ref{sec:experiment} when the length of $T_{pack}$ is long enough then most of the interrupts are generated by $T_{abs}$. When this occurs the variance in the inter-arrival timing between measurements decreases, which leads to less information being contained in the inter-arrival timings. This situation is similar to uniform sampling and more relevant information is contained in the number of packets per measurement. Thus, although we designed our detection method using knowledge of the measurement system, and our method outperforms PAD under typical measurement conditions there are still situations were another method, like PAD, will be a better choice.  This highlights the fact that we set out to prove initially, that careful consideration must be taken when analyzing measurements that were obtained from a software-based measurement system. Figure~\ref{fig:graph} shows the relationship between the information contained in the inter-arrival timings versus the information contained in the number of packets per measurement. The line corresponds to a fixed selection of measurement system parameters, and as indicated on the graph if the input traffic rate increases then the signal information will slide from having more relevant information in the measurement inter-arrivals (good for PDMM) to the information being contained in the number of packets per measurement (good for PAD).

\bibliographystyle{IEEEbib}
\bibliography{InternalReportFinal}

\end{document}